\documentclass[authoryear,preprint,review,12pt]{elsarticle}

\usepackage{amssymb}

\usepackage{lineno}

\journal{Planetary and Space Science}

\begin{document}

\begin{frontmatter}


 \title{The shape and structure of small asteroids as a result of sub-catastrophic collisions}
 

\author[label1]{Martin Jutzi}
\address[label1]{Physics Institute, University of Bern, NCCR PlanetS, Gesellschaftsstrasse 6, 3012 Bern, Switzerland}
\ead{martin.jutzi@space.unibe.ch}




\begin{abstract}

The overall shape, internal structure and surface morphology of small bodies such as asteroids and comets are determined to a large degree by the last global-scale impact or disruption event. Depending on the specific energy, impacts lead to a large spectrum of outcomes.
Sub-catastrophic disruptions take place in an energy range between cratering impacts and catastrophic disruptions. Although less energetic than catastrophic events, they can still significantly alter the overall shape and structure of the target body. This has been demonstrated recently in the case of bi-lobe cometary nuclei \citep{Jutzi:2017jb}. 
Here we present results of a subsequent study on the shapes of asteroids resulting from such collisions. Sizes ranging from a few hundred meters to a few kilometers are considered.

We show that impacts on elongated rotating asteroids frequently lead to the formation of contact binaries. Our results confirm that this mechanism is robust and works for a large range of asteroid sizes and impact velocities. Scaling-laws for the prediction of the size and velocity dependent specific energies required for successful bi-lobe formation are presented. Based on these scaling laws, the expected frequency of such sub-catastrophic impacts is calculated and is compared to the one of catastrophic disruptions, which require much higher specific energies and are more rare. 

Our analysis suggest that the shapes and structures of a large fraction of small asteroids as observed today may be the result of the last major sub-catastrophic impact.
 
\end{abstract}


\end{frontmatter}


\newpage

\section{Introduction and Motivation}\label{s:intro}
The objects of the Solar System's small body populations - leftovers from the planet formation process - are a result of a billion-year-long collisional evolution of varying intensity, even if they were born big \citep[e.g.][]{Morbidelli:2009dd}. Their current shape, structure and composition provide important clues to their origin and history, and therefore represent a window to the early stages of the formation of the Solar System. In the asteroid population, the observed asteroid families \citep[e.g.][]{Farinella:1996fd} are direct evidence for such a collisional evolution. Asteroids smaller than about 50 km in diameter are the result of a break-up of a larger parent body \citep{Bottke:2005bd}. Their properties such as their shapes, interior structure and spin state are determined to a large degree by the last major (global scale) impact event. Important questions are how much asteroids were processed by past collisions and to what extend they retained a record of processes that took place during the formation and early evolution of the Solar System. 
Depending on the specific energy, collisions lead to a large spectrum of outcomes, ranging from small impact craters to the complete disintegration of the target, accompanied by significant loss of mass (equal to half of the initial mass in so-called catastrophic disruptions). The average impact velocity in today's asteroid belt is around 5 km/s \citep{Bottke:1994bn}, which means that low velocity collisions, leading to merging of the the involved objects, are extremely rare. Therefore, direct mergers of similar-sized bodies, leading to bi-lobe shapes \citep{Jutzi:2015ja,Sugiura:2018sk} or flattened objects with equatorial ridges \citep{Leleu:2018lj} are unlikely. The vast majority of impacts lead to disruption and erosion.

Recent progress in numerical modelling has allowed for the simulation of the entire process, the collisional disruption followed by the gravitational re-accumulation, and to compute the final shapes and structures of the objects resulting from an asteroid collision. Studies of catastrophic disruptions predict a large spectrum of shapes and properties of the resulting fragments \citep[e.g.][and Walsh et al., 2018, submitted; Michel et al. 2018, submitted]{Michel:2013mr,Schwartz:2018sm,Sugiura:2018sk}. 

\begin{figure}
\begin{center}
\includegraphics[width=14cm]{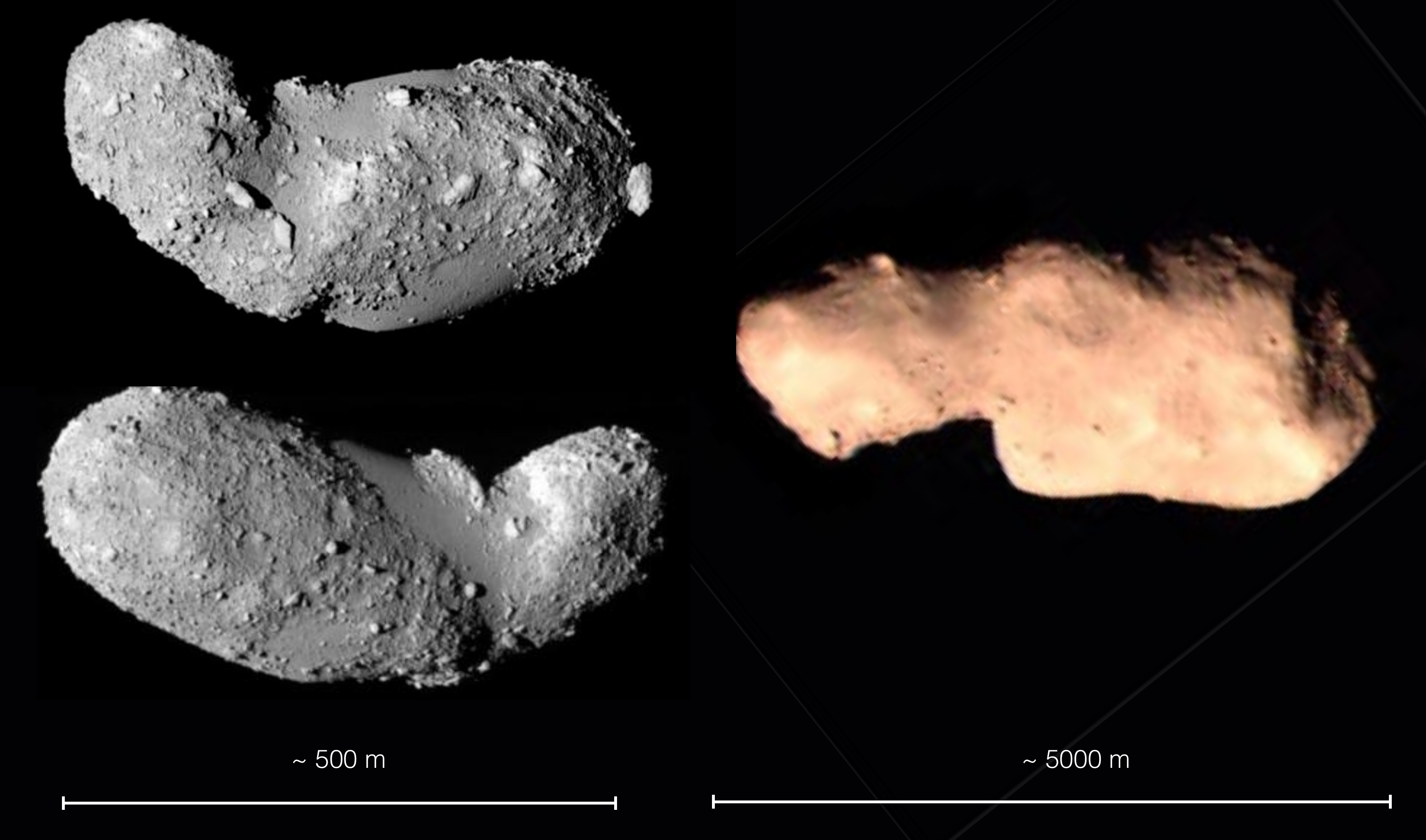}\\
\caption{Examples Asteroid contact binaries at different scales. Shown are asteroid Itokawa (left) observed by JAXA's Hayabusa spacecraft and asteroid Toutatis (right) imaged by the Chinese lunar probe Chang'e 2.}
\label{fig:examples}
\end{center}
\end{figure}

On the other hand, smaller-scale, sub-catastrophic collisions can also cause an alteration of the overall structure  \citep[e.g.][]{Jutzi:2017jb}. Such 'shape-changing' collisions are in a energy range between cratering and catastrophic disruption. Their frequency can be orders of magnitude higher than catastrophic events \citep{Jutzi:2017jb}, depending on the slope of the size-distribution of the impactor population. For asteroids smaller than $\sim$ 100 m in diameter, the slope of the impacting population is steep. Collisional evolution studies found the \emph{cumulative} slope to be in the range of -2.75 \citep{Obrien:2005og} and -2.65 \citep{Bottke:2005bd}. Such impacting populations are also relevant for the collisional evolution of current NEAs such as Itokawa \citep{Michel:2009mo}. An impact population with a steep size distribution leads to a collision regime dominated by small impacts. The majority of shapes and structures of small asteroids as observed today may therefore be the result of the last major sub-catastrophic collision, involving specific energies far below what is required for a catastrophic disruption. We analyze this in the case of contact binaries, considering asteroid Itokawa as an example, extending the work by  \citet{Jutzi:2017jb} - which was focussed on cometary nuclei - to asteroids. Contact binaries are frequent among the asteroid population (Mainbelt and NEO). Two examples are shown in Figure \ref{fig:examples}.

 \citet{Jutzi:2017jb}  showed that bi-lobed cometary nuclei result frequently from sub-catastrophic impacts into rotating, elongated, porous bodies. The goal of this study is to investigate whether or not this mechanism also works in the case asteroids, which, even if they are porous, generally have a higher density and material strength than cometary nuclei. We also want to study at which scales and impact velocity range bi-lobed shapes can be produced by this type of sub-catastrophic impacts (does it work at very small scales as well?), and what is the resulting interior structure and surface properties. Finally, the frequency of sub-catastrophic collisions is  compared to the much more energetic catastrophic disruptions to determine the respective probability of such events to form a bi-lobe type structure. 

Our model approach is presented in section \ref{s:methods}. The results in terms of bi-lobe shape formation  are presented in section \ref{s:results} and the size- and velocity-dependence discussed in section \ref{s:scaling}. The frequency and probabilities of the events are analyzed in section \ref{s:frequency}. Conclusions and outlook are given in section \ref{s:conclusions}. 

\section{Model approach}\label{s:methods}
In this section we describe our modeling approach as well as the assumptions regarding the asteroid properties and initial conditions. 

\subsection{Shock-physics code}\label{s:shockphysicscode}
We use a Smoothed Particle Hydrodynamics (SPH) impact code \citep{Benz:1994ij,Benz:1995hx,Nyffeler:2004tz,Jutzi:2008kp,Jutzi:2015gb} that includes self-gravity as well as material strength models. To model fractured, granular material, a pressure-dependent shear strength (friction) is included by using a standard Drucker-Prager yield criterion \citep{Jutzi:2015gb}. Porosity is modeled using a sub-resolution approach based on the P-alpha model  \citep{Jutzi:2008kp}, taking into account the enhanced dissipation of energy during compaction of porous materials. We use a quadratic crushcurve with $P_e$ = 1 10$^6$ dyn/cm$^3$ and  $P_s$ = 1 10$^8$ dyn/cm$^3$ \citep{Jutzi:2008kp}.

We further use the Tillotson Equation of State (EOS) with parameters for basalt as given in \citet{Jutzi:2009ht} but use a reference density of $\rho_0$ = 3 g/cm$^3$. We also adapt the bulk modulus to $A$ = 2.7 10$^9$ dyn/cm$^3$ resulting in an elastic wave speed of the order of 300 m/s, consistent with the assumption of a rubble-pile structure (see next section).  

\subsection{Model of the internal structure}
The target asteroids used in this study are assumed to have a rubble-pile like structure, which means that the SPH particles are damaged but they experience friction according to the pressure-dependent strength model. The overall initial porosity is 40 $\%$. 
The scale of porosity is defined in comparison with the other relevant dimensions involved in the problem, such as the size of the projectile and/or crater. Using a sub-resolution P-alpha type porosity model implicitly assumes that the scale of the porosity is smaller than the scale of the impactors (i.e., in the case of the $\sim$ km-sized asteroids used in our study, the scale of porosity is assumed to be smaller than $\sim$ 10 meters). 
A body containing such small-scale porosity may be crushable: cratering on a microporous asteroid is an event involving compaction rather than ejection \citep{Housen:1999iz}.

\subsection{Initial conditions}\label{sec:inicond}

We consider two different target sizes (with radii $R_t$ = 500 m and $R_t$ = 2500 m) and a range of impact velocities (100 m/s to 3000 m/s) and impactor sizes. Due to numerical limitations (resolvable projectile size), the velocities are in the lower range of the asteroid impact velocity distribution \citep{Bottke:1994bn}. 
However, scaling laws allow for an extrapolation of the results to higher velocities (section \ref{s:scaling}).

The initial targets are have ellipsoidal shapes with axis ratios ranging from 0.4 to 0.6 and initial rotation periods of 6 hours, properties that are typical for asteroids \citep{Cibulkova:2017cn}. 

The range of impact geometries and target rotation states used in this study are guided by our previous work on bi-lobe comet formation \citep{Jutzi:2017jb}. A more complete exploration of the huge parameter space of possible impact conditions is left for subsequent studies. 

\section{Results}\label{s:results}
\subsection{Example of contact binary formation}
\begin{figure}
\begin{center}
\includegraphics[width=14cm]{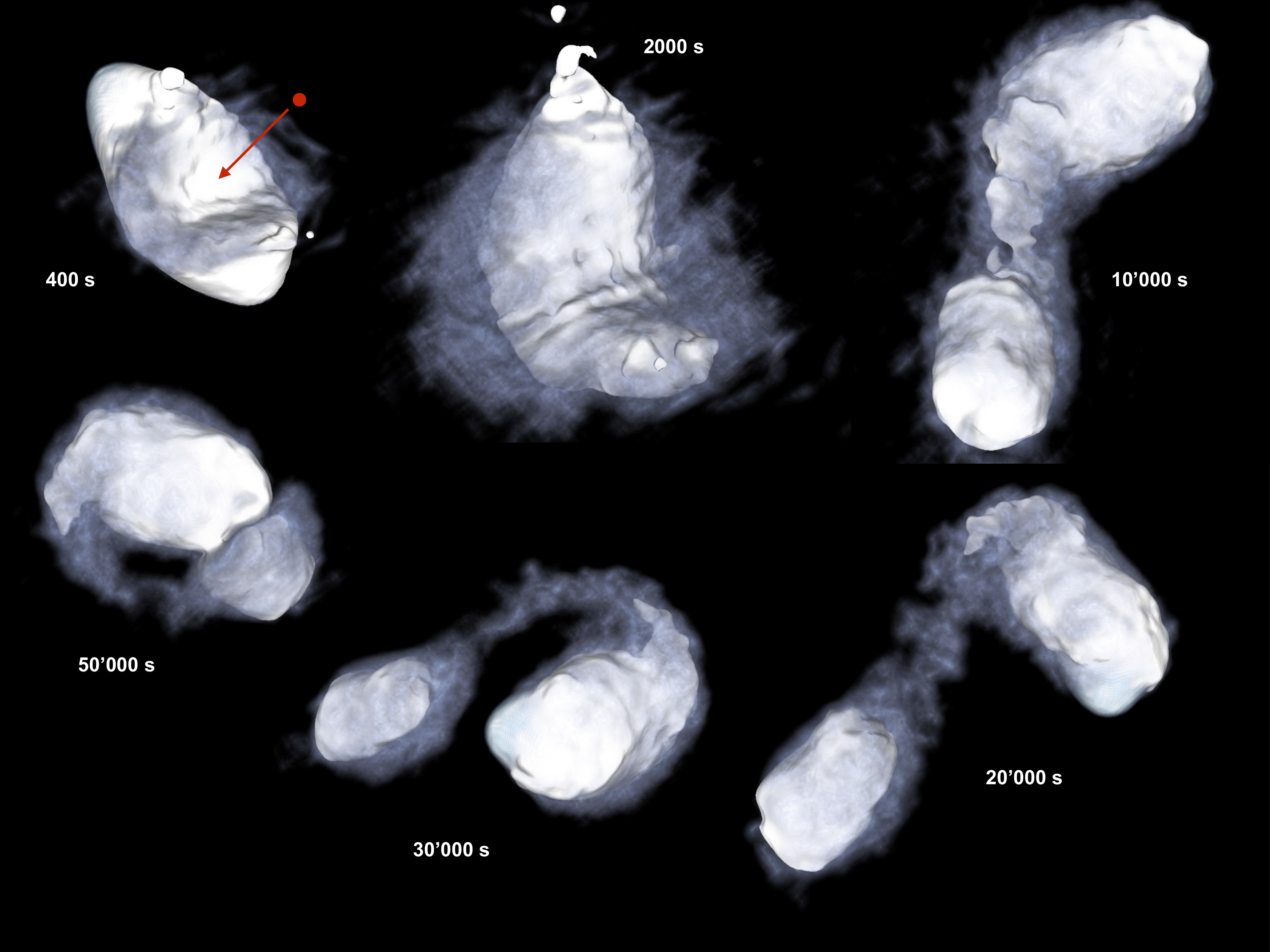}
\caption{Time sequence of a shape-changing sub-catastrophic collision resulting in a contact-binary. Shown are snapshots of a 3D SPH simulation. The target has size of $R_t$ = 500 m, an axis ratio of 0.4 and an initial rotation period of 6 h. The projectile has a radius of 6 m and an impact velocity of 2300 m/s.}
\label{fig:sim_example}
\end{center}
\end{figure}

An example of a sub-catastrophic collision leading to a final body with a bi-lobe structure is shown in Figure \ref{fig:sim_example}. As found by \citet{Jutzi:2017jb} in the case of cometary nuclei, the central impact into the elliptical target leads to a splitting into two components, which is accompanied by re-accumulation of ejected material. The two resulting objects are gravitationally bound and merge within 1-2 days to form a contact binary. 

\subsection{Results for various initial conditions}
\begin{figure}
\begin{center}
\includegraphics[width=14cm]{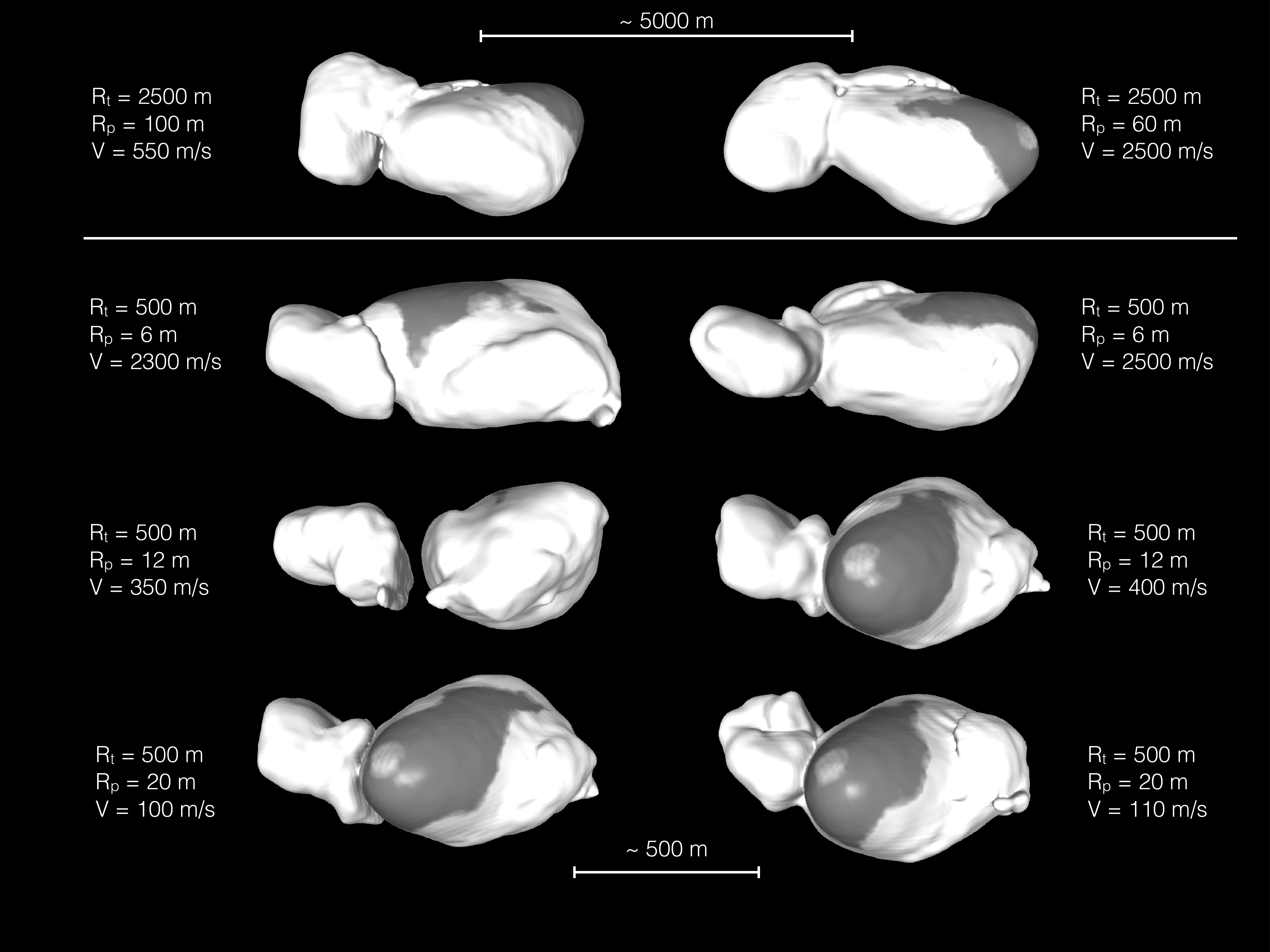}
\caption{Final shapes resulting from sub-catastrophic impacts with different initial conditions.  Shown are iso-density surfaces with colors indicating regions with (white) and without (gray) ejecta coverage. The targets have an axis ratio of 0.4 and an initial rotation period of 6 h.}

\label{fig:sim_various_0.4}
\end{center}
\end{figure}

\begin{figure}
\begin{center}
\includegraphics[width=14cm]{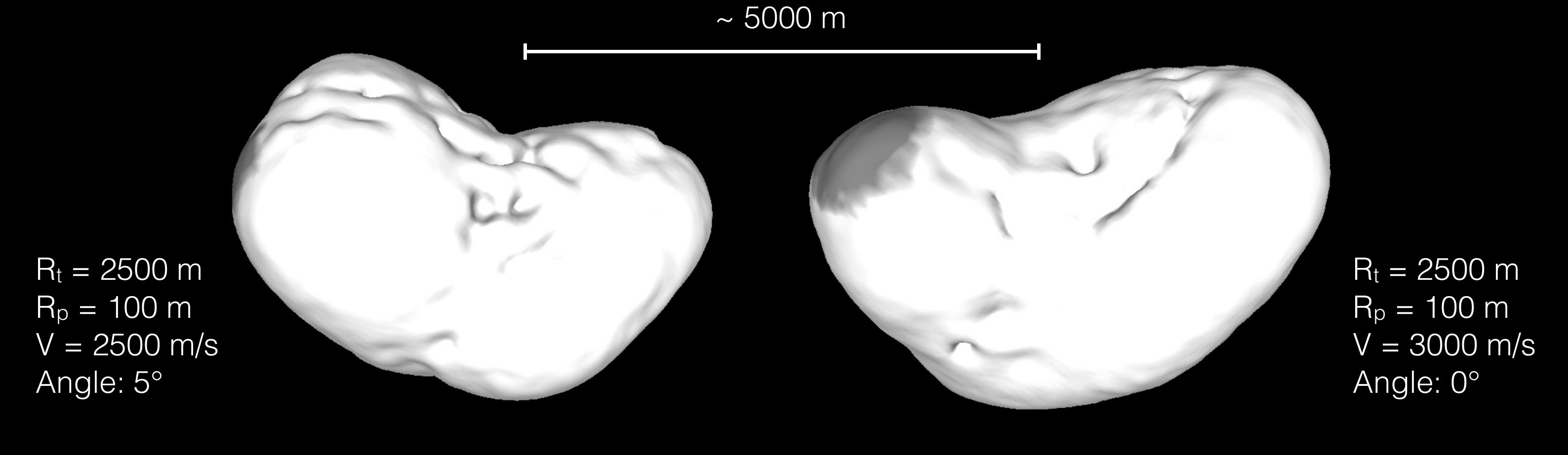}
\caption{Same as Figure \ref{fig:sim_various_0.4} but for targets with an axis ratio 0.6.}

\label{fig:sim_various_0.6}
\end{center}
\end{figure}

In Figures \ref{fig:sim_various_0.4} and \ref{fig:sim_various_0.6}, successful cases of bi-lobe formation are shown for various initial conditions (section \ref{sec:inicond}). In a post-processing analysis, material that was ejected and later re-accumulated onto the individual lobes is identified\footnote{Ejected SPH particles are defined as those which have a ratio of $h/h_0$ $>$ $\sim$ 1.3 at t = 5000 s, where $h_0$ is the initial smoothing length and $h$ is the current one.} and is shown in white. On the other hand, dark colors show little processed original surface.
 
Our model of the formation of bi-lobe asteroids generally produces two lobes with a relatively coherent interior (as illustrated in Figure 9 of  \citet{Jutzi:2017jb}),  covered by re-accumulated material. However, on the final bi-lobe body some portions of the original surface are still visible. The distinct areas of boulders and smooth surfaces observed on asteroid Itokawa show similarities to the patterns produced by the ejecta re-accumulation in the simulations. A formation of Itokawa by the bi-lobe forming sub-catastrophic impacts considered here may also be consistent with evidence of a recent global reshaping event \citep{Tatsumi:2018ts}.  We note that the details of the outcomes of our simulations strongly depend on the rotation state and impact conditions.

\section{Size and velocity scaling}\label{s:scaling}
\begin{figure}
\begin{center}
\includegraphics[width=14cm]{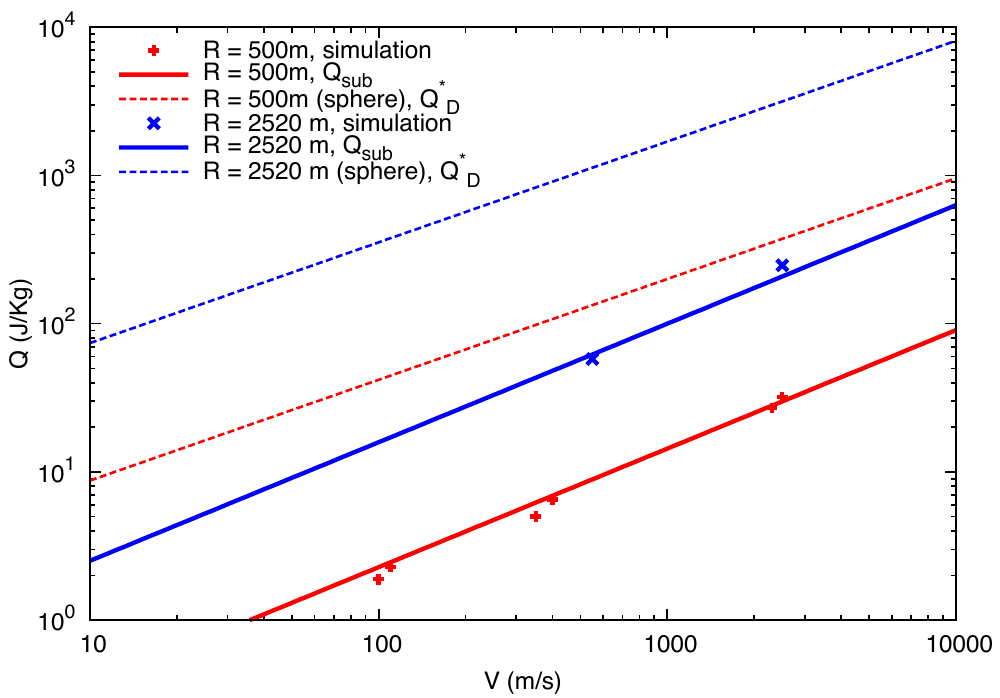}
\caption{Specific energies of successful bi-lobe forming impact simulations (as displayed in Figs. \ref{fig:sim_various_0.4}). Also shown is the scaling law  $Q_{sub} = a R^{3\mu} V^{2-3\mu}$  using a fixed set of parameters a = 0.4 and u = 3.3 10$^{-5}$. For comparison, the scaling law $Q^*_D$ for catastrophic disruptions  of spherical, non-rotating bodies based on data from \citet{Jutzi:2015gb} and  \citet{Jutzi:2019jm} is shown as well.}
\label{fig:scaling}
\end{center}
\end{figure}

The specific impact energies of the successful cases of bi-lobe formation (for axis ratio 0.4) are plotted in Figure \ref{fig:scaling} for two different target sizes as a function of velocity. As in  \citet{Jutzi:2017jb} we use a scaling law 
\begin{equation}\label{eq:scaling}
Q_{sub} = a R^{3\mu} V^{2-3\mu}
\end{equation}
to compare with the obtained results. While \citet{Jutzi:2017jb} have already shown that the velocity-dependence can be well reproduced, the results presented here also allow for a test of the size-dependence. As shown in Figure \ref{fig:scaling}, for a given target size and impact velocity, the specific impact energy required for successful bi-lobe forming impact can be well predicted by the scaling law (equation \ref{eq:scaling}) using the fixed parameters $a$ = 3.3 10$^{-5}$ and $\mu$ = 0.4. This suggests that this bi-lobe formation mechanism is robust and works over a large range of scales and velocities.

As also shown in Figure \ref{fig:scaling}, the specific impact energies $Q^*_D$ required for catastrophic disruptions  of spherical, non-rotating bodies of the same size is about an order of magnitude higher than $Q_{sub}$ required for bi-lobe formation. For this comparison we use $Q^*_D$ curves which are based on  asteroid disruption calculations performed in \citet{Jutzi:2015gb} and  \citet{Jutzi:2019jm}, which used targets with similar properties.

\section{Collision frequencies and probabilities}\label{s:frequency}
In this section we compare the frequency of the sub-catastrophic collisions considered here with the one of the much more energetic catastrophic disruptions, which also has been suggested to possibly form bi-lobe structures \citep[e.g.][]{Schwartz:2018sm} and Itokawa-like shapes \citep{Michel:2013mr,Mazrouei:2014mb}. 

As in \citet{Jutzi:2017jb}, the number of impacts on a target of size $R_t$ by projectiles within the size range $R_{min} \leq R_t \leq R_{max}$ during a time interval $\delta t$ is written as

\begin{equation}\label{eq:nbcol}
N = P_i \delta t \int^{R_{max}}_{R_{min}} \pi (R_t+R_p)^2 N_p(R_p)dR_p
\end{equation}

where we use

\begin{equation}
R_{min} = (2 Q_{min} V)^{1/3} R_t / V
\end{equation}

and

\begin{equation}
R_{max} = (2 Q_{max} V)^{1/3} R_t / V
\end{equation}

We further consider the range of
\begin{equation}
Q_{max} = 1.5 Q_{crit}
\end{equation}

\begin{equation}
Q_{min} = 0.5 Q_{crit}
\end{equation}
 
The critical specific energy is obtained by using the scaling law
 
\begin{equation}
Q_{crit} = a R^{3\mu} V^{2-3\mu}
\end{equation}

where we use the parameters $a$ = 3.3 10$^{-5}$ and $\mu$ = 0.4 obtained in section \ref{s:scaling} for the sub-catastrophic collisions ($Q_{crit}=Q_{sub}$) and $a$ = 5 10$^{-4}$ and $\mu$ = 0.44 for catastrophic disruptions ($Q_{crit}=Q^*_D$). As shown in Fig.  \ref{fig:scaling},  $Q^*_D$ is about an order of magnitude higher than $Q_{sub}$ for the same target size. In this example calculation, we consider sizes of the original target of  $R_t$ = 500 m in the case of sub-catastrophic collisions and  $R_t$ = 1000m in the case of catastrophic disruptions (where the bi-lobe object would form from a small part of the larger parent body as a re-accumulated fragment).

In oder to compute the ratio of the number of sub-catastrophic collisions to the number of catastrophic disruptions $N_{sub}/N_{cat}$ we use equation (\ref{eq:nbcol}) in each case and further assume a differential size distribution characterized by the exponent $q$ 
\begin{equation}
\frac{dN}{dr}  \sim r^q
\end{equation}

The ratio $N_{sub}/N_{cat}$ is independent of $P_i$ and $dt$ and is given in Table \ref{table:numberofcollisions} for different slopes $q$ and an impact velocity of 5 km/s. We further multiply the ratio $N_{sub}/N_{cat}$ by the number ratio of available targets $N_{R,500m}$/$N_{R,1000m}$ to obtain an estimate of the ratio of the number of events $N_{sub'}/N_{cat'}$. The obtained number ratios strongly depend on the exponent $q$ of the size distribution. For the impactor sizes considered here, the impacting population has a steep slope ($q$ = -3 to -4; see section \ref{s:intro}) and $N_{sub'}/N_{cat'}$ becomes very large (Table \ref{table:numberofcollisions}), leading to a collision regime dominated by sub-catastrophic impacts.

\begin{table*}
\caption{Exponent of differential size distribution $q$, ratio of number of impacts (on a given target) $N_{shape}/N_{cat}$, ratio of number of available targets $N_{R,500m}$/$N_{R,1000m}$, ratio of number of events $N_{sub'}/N_{cat'}$. }    
\label{table:numberofcollisions}      
\centering                        

\begin{tabular}{c c c c }        
\hline\hline            
  $q$ & $N_{sub}/N_{cat}$ & $N_{R,500m}$/$N_{R,1000m}$ & $N_{sub'}/N_{cat'}$ \\
 \hline        
-2 & 1.4 & 4.0 & 5.8 \\
-3 & 8.8 & 8.0 & 71 \\
-4 & 54 & 16 & 865  \\
\hline                               
\end{tabular}
\end{table*}

\section{Conclusions and outlook}\label{s:conclusions}
Our results show that sub-catastrophic collisions - with specific energies in the range between cratering impacts and catastrophic disruptions - can cause an overall change of the shape and structure of asteroids. As it is also demonstrated here, such events are much more frequent than the more energetic catastrophic disruptions, which suggests that  a large fraction of the current shapes and interior structures of small asteroids are determined to a large degree by the last sub-catastrophic impact. We show that the contact binary asteroid Itokawa may indeed have resulted from such an event.

Observed asteroids are not spherical but typically have some elongation and are rotating. The results presented here show that these properties enhance the degree of shape change and enable the formation of bi-lobe objects by sub-catastrophic impacts. In studies of the collisional formation and evolution of small bodies it is therefore crucial to take into account non-spherical target characteristics and rotation. However, only a limited range of initial conditions have been investigated and a huge parameter space remains to be explored in future studies. For instance, only slightly different (more energetic) initial conditions than the ones considered here may lead to two unbound components, possibly with characteristics similar to the observed asteroid pairs \citep{Pravec:2019pf}.

\section*{Acknowledgments}
M.J. acknowledges support from the Swiss National Centre of Competence in Research PlanetS. 



\bibliography{bibdata.bib}



\newpage

\end{document}